\documentclass[aps,prl,reprint,superscriptaddress]{revtex4-2}

\usepackage{braket}
\usepackage{graphicx}
\usepackage{amsmath}
\usepackage{amssymb}
\usepackage{xcolor}
\usepackage{xurl}
\usepackage[
    colorlinks = true,
    urlcolor   = blue,
    linkcolor  = red,
    citecolor  = green
]{hyperref}

\begin{document}

\title{Dissipative phase transitions in quantum reservoir computing}

\author{Da Zhang}
    \affiliation{Center for Quantum Technology Research and Key Laboratory of Advanced Optoelectronic Quantum Architecture and Measurements (MOE), \\ School of Physics, Beijing Institute of Technology, Beijing 100081, China}
  
\author{Zhang-Qi Yin}
    \email{zqyin@bit.edu.cn}
    \affiliation{Center for Quantum Technology Research and Key Laboratory of Advanced Optoelectronic Quantum Architecture and Measurements (MOE), \\ School of Physics, Beijing Institute of Technology, Beijing 100081, China}

\date{\today}

\begin{abstract}
Enhanced performance of quantum reservoir computing has been associated with dynamical phase transitions, but whether this connection extends to dissipative systems and which relaxation mechanisms underlie it remain insufficiently understood. We systematically compare driven-dissipative Kerr reservoirs across first-order and continuous dissipative phase transitions and find enhanced memory and nonlinear processing near both phase boundaries. Although the closure of the Liouvillian gap marks the transitions in the thermodynamic limit, computational performance does not generally follow gap suppression. An exact post-training Liouvillian-mode decomposition quantitatively attributes the trained memory to intrinsic relaxation channels. It shows that the gap mode contributes only weakly, while finite-rate modes and their cross-contributions dominate the enhanced capacity. These results go beyond phenomenological correlations by directly linking memory capacity to the intrinsic Liouvillian relaxation spectrum. Moreover, these findings provide a physical basis for designing dissipative quantum reservoirs and testing memory-enhancement mechanisms in experimentally accessible Kerr platforms.
\end{abstract}

\maketitle

\textit{Introduction.—}
Dissipation is usually regarded as detrimental to quantum information processing. However, when properly controlled, it can also serve as a resource for quantum state preparation and stabilization~\cite{barreiro2011quantumsimulation,lin2013dissipativestate,shankar2013dissipativestate,2018discorrection}, quantum error correction~\cite{reiter2017dissipativecorrection}, and open-system quantum simulation~\cite{barreiro2011quantumsimulation,so2024dissimulation}. This perspective is particularly relevant to quantum reservoir computing (QRC), where a fixed quantum dynamical system processes a temporal input stream and only the classical readout is trained~\cite{Fujii2017,PRXQuantumqrcgen,tapioqrcgen,guoqrcgen,zia2025qrcgen,qrcgenspectral,houqrcgen,qrcoscillator,PRXQuantumqrcgenxun,qrcoscijc}. For time-series processing, the reservoir must retain recent inputs while progressively erasing the remote past. Open-system reservoirs can naturally acquire this fading-memory property from their intrinsic relaxation dynamics, without requiring repeated state resets~\cite{sannia2024dissipationqrc,qrcdisearly,qrcdisdamp,qrrqrcnoise23,qrcnoise}. The central question is whether open quantum dynamics can exhibit a computational edge near a dissipative phase boundary, and which features of relaxation underlie the resulting enhancement in information processing.

Related computational optima have been reported in closed-system QRC near transitions between localized and ergodic dynamics~\cite{PhysRevLettqrcphase,Xia2022qec,our} and near the onset of many-body quantum chaos~\cite{qrcphseedgeofchaos}. These studies establish a phenomenological connection between dynamical boundaries and enhanced QRC performance. In dissipative systems, this question can be addressed through the Liouvillian spectrum. The Liouvillian gap sets the slowest relaxation rate, and its closure marks a dissipative phase transition~\cite{mori2020resolvinggap,ciutispectral}. The same relaxation dynamics also governs how past information is erased from the reservoir~\cite{sannia2024dissipationqrc}. This provides a direct route for relating dissipative criticality to reservoir memory through the relaxation spectrum itself.

Dissipative phase transitions can arise through different physical mechanisms and exhibit distinct spectral structures~\cite{disphaselukin,ciutispectral}. We consider two representative driven-dissipative Kerr reservoirs. In the linearly driven reservoir, a first-order transition arises from switching between coexisting phases~\cite{Drummond19801st,Casteels171st}. Its soft mode becomes slow only near the transition~\cite{Casteels171st,ciutispectral}. In the quadratically (two-photon) driven Kerr reservoir, a continuous transition is associated with $Z_2$ symmetry breaking~\cite{ciutiphase,ciutispectral}. The gap remains closed throughout the symmetry-broken phase~\cite{ciutispectral}. These different spectra imply different patterns of information decay and may therefore lead to different reservoir behavior. This contrast allows us to examine how the transition order shapes the mechanism underlying the performance enhancement and whether the trained memory is carried mainly by the gap mode or by a broader set of low-lying relaxation modes.

Here, we show that both reservoirs exhibit performance optima near their respective transitions, yet with distinct relations to their Liouvillian spectrum. Information-processing capacity and nonlinear autoregressive moving-average benchmarks~\cite{Fujii2017,Ghosh2019QuantumReservoirProcessing,Martinezqrc2023Information,Dambre2012IPC,carles26qrccqed} show that these optima closely track the spectral reorganization. To uncover the origin of this correspondence, we develop an exact post-training, mode-resolved decomposition that identifies the contributing relaxation channels. The decomposition shows that the gap mode contributes only weakly, whereas finite-rate modes and their cross-contributions dominate the trained capacity, identifying finite-rate spectral reorganization rather than the slowest scale alone as the origin of the enhancement. We further show that the relevant Liouvillian spectral features can be extracted from measurable dynamical correlations, demonstrating the experimental feasibility of our framework and providing an accessible diagnostic for guiding reservoir operation.

\begin{figure}[htp]
    \centering
    \includegraphics[width=\columnwidth]{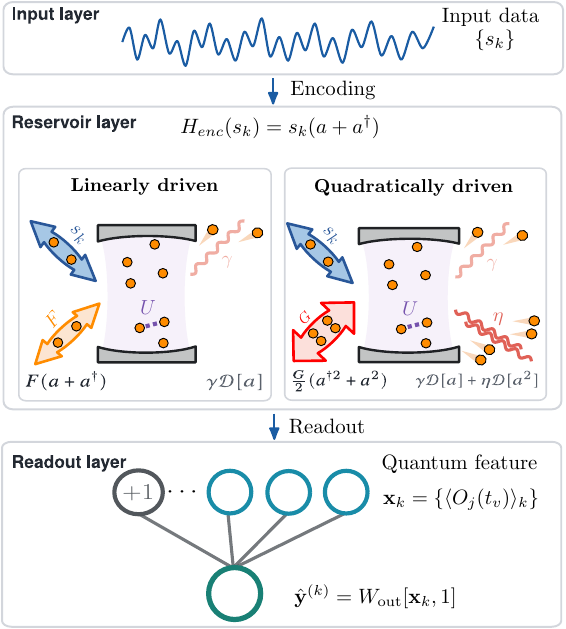}
    \caption{Schematic of the dissipative-QRC protocol. A scalar input ${s_k}$ is encoded through the quadrature drive $H_{\rm enc}=s_k(a+a^\dagger)$ and processed by Kerr reservoirs exhibiting either a first-order or continuous ($Z_2$)-symmetry-breaking dissipative phase transition. Virtual-time measurements form the feature vector $\mathbf{x}_k$, and only the affine readout $\hat y^{(k)}=W_{\rm out}[\mathbf{x}_k,1]^T$ is trained.
}
    \label{fig:workflow}
\end{figure}

\textit{Driven-dissipative Kerr reservoirs.—}
To distinguish universal signatures of dissipative criticality from transition-specific effects, we consider two single-mode Kerr reservoirs realizing qualitatively different dissipative phase transitions, one exhibiting a first-order transition~\cite{Casteels171st} and the other a continuous ($Z_2$)-symmetry-breaking transition~\cite{ciutiphase,ciutispectral}. Their QRC implementation is summarized in Fig.~\ref{fig:workflow}. Despite their distinct critical behavior, both reservoirs can be described within a common open-system framework, with zero-input dynamics governed by
\begin{equation}
    \mathcal{L}_{0}\rho =-i[H_0,\rho] +\frac{\gamma}{2}\mathcal{D}[a]\rho +\frac{\eta}{2}\mathcal{D}[a^2]\rho ,
    \label{eq:main-zero-input-liouvillian}
\end{equation}
where $\mathcal{D}[O]\rho =2O\rho O^\dagger-\{O^\dagger O,\rho\}$, and $H_0$ is specified below for the two reservoirs.

The choices of $H_0$ and $\eta$ distinguish the two reservoirs. For the linearly driven reservoir, $H_0=H_{\rm lin}$, with
\begin{equation}
    H_{\rm lin} =-\Delta a^\dagger a +\frac{U}{2}a^{\dagger 2}a^2+F(a+a^\dagger),
    \label{eq:main-first-model}
\end{equation}
and $\eta=0$~\cite{Casteels171st}. Here $\Delta$ is the pump-cavity detuning, $U$ is the Kerr nonlinearity, and $F$ is the coherent-drive amplitude. Under the thermodynamic scaling $U=\tilde U\gamma/N$ and $F=\tilde{F}\sqrt{N}\gamma$, the rescaled coherent drive $\tilde F$ tunes a first-order dissipative phase transition~\cite{Drummond19801st,Casteels171st,Chen20231stexp}. At finite $N$, the steady state remains unique, while the transition region is characterized by long-lived switching between competing metastable configurations.

For the quadratically driven reservoir, we instead use a two-photon-driven Kerr oscillator with $H_0=H_{\rm quad}$,
\begin{equation}
    H_{\rm quad}=-\Delta a^\dagger a+\frac{U}{2}a^{\dagger 2}a^2+\frac{G}{2}\left(a^{\dagger 2}+a^2\right).
    \label{eq:main-second-model}
\end{equation}
Under the scaling $U=\tilde U\gamma/N$, $\eta=\tilde\eta\gamma/N$, and $G=\tilde G\gamma$, the rescaled two-photon drive $\tilde G$ tunes a continuous $Z_2$-symmetry-breaking transition~\cite{ciutiphase,ciutispectral,beaulieu2025exp12}.

To use these dissipative systems as temporal reservoirs, the input sequence $\{s_k\}$ is encoded as a piecewise-constant quadrature drive $H_{\rm enc}=s_kX$, with $X=a+a^\dagger$. During the $k$th interval of duration $\Delta t$, the reservoir evolves without reset under
\begin{equation}
    \mathcal{L}_{s_k}\rho=\mathcal{L}_0\rho-is_k[X,\rho].
    \label{eq:main-input-liouvillian}
\end{equation}
The linearly driven reservoir experiences a modulation of its coherent drive, whereas in the quadratically driven reservoir the input is a parity-odd field that directly accesses the symmetry-breaking sector. Because the state is inherited between intervals, the input-to-state map retains history: after $s_k$ changes, the reservoir is generally not stationary under the new generator, so transients from earlier inputs persist.

Each interval is sampled at $V$ virtual nodes. The observables $\mathcal{O}=\{n,X,P,X^2,P^2,(XP+PX)/2\}$ form the feature vector $\mathbf{x}_k=\{\operatorname{Tr}(O_j\rho_{k,v})\}_{j,v}$. Virtual-node sampling resolves distinct intra-interval trajectories of modes with similar terminal amplitudes. The target is reconstructed by an affine readout $\hat y^{(k)}=W_{\rm out}[\mathbf{x}_k,1]^T$, with only $W_{\rm out}$ trained by ridge regression. The nonlinear fading-memory representation is therefore generated by the driven-dissipative dynamics, while the readout selects its task-relevant components. Numerical details are given in the Supplemental Material~\cite{SM}.

\begin{figure}[htp]
    \centering
    \includegraphics[width=\columnwidth]{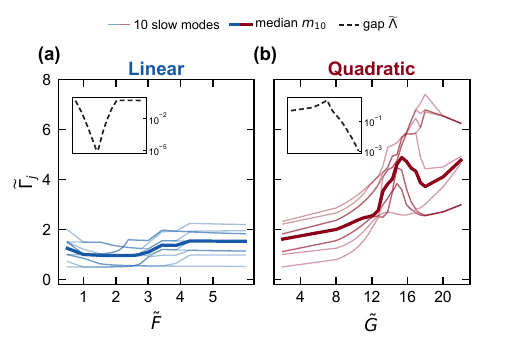}
    \caption{
    Liouvillian spectral reorganization across (a) the first-order and (b) the continuous dissipative phase transition. The ten slowest nonstationary decay rates beyond the gap are shown individually, with their median $m_{10}$ highlighted by the bold curve. Insets show the Liouvillian gap $\widetilde{\Lambda}\equiv\Lambda/\gamma$ on a logarithmic scale.}
    \label{fig:spectral_reorganization}
\end{figure}

\textit{Liouvillian spectral reorganization.—}
We first examine how the two dissipative phase transitions reorganize the intrinsic relaxation dynamics of the reservoir. The relevant decay channels are defined by the zero-input Liouvillian,
\begin{equation}
    \mathcal{L}_{0}(\zeta,N)R_\mu = \lambda_\mu(\zeta,N)R_\mu ,
    \label{eq:main-liouvillian-spectrum}
\end{equation}
where $\zeta=\tilde F$ for the linearly driven reservoir and $\zeta=\tilde G$ for the quadratically driven reservoir. The eigenvalue $\lambda_0=0$ corresponds to the stationary state, while $\operatorname{Re}\lambda_\mu<0$ for nonstationary modes. The slowest nonstationary relaxation is characterized by the Liouvillian gap,
\begin{equation}
    \Lambda(\zeta,N) = \min_{\mu\neq0} \left[-\operatorname{Re}\lambda_\mu(\zeta,N)\right].
    \label{eq:main-liouvillian-gap}
\end{equation}

For the dynamics generated by $\mathcal{L}_{0}$, a mode with decay rate $\Gamma_\mu=-\operatorname{Re}\lambda_\mu$ is attenuated over a time interval $\Delta t$ by $e^{-\Gamma_\mu\Delta t}$. The zero-input spectrum therefore defines the intrinsic relaxation timescales of the reservoir and provides a physically distinguished spectral reference for characterizing their reorganization across the dissipative phase transition. The gap isolates the longest asymptotic relaxation timescale, whereas the remaining low-lying modes describe finite-rate relaxation on shorter timescales. A phase transition can therefore modify the surrounding low-lying spectrum in a manner that is not captured by the gap alone. To characterize this spectral reorganization, we assign each mode the dimensionless decay rate $\widetilde{\Gamma}_{\mu}\equiv -\operatorname{Re}(\lambda_{\mu}/\gamma)$. We then track the ten slowest nonstationary modes beyond the gap and summarize their collective evolution by the median decay rate $m_{10}$. The individual rates together with $m_{10}$ distinguish softening confined to the gap mode from a broader reorganization of the surrounding low-lying spectrum.

Figure~\ref{fig:spectral_reorganization} reveals qualitatively distinct reorganizations of the low-lying spectrum across the two transitions. For the linearly driven reservoir, the sharp gap minimum near coexistence is accompanied by localized slowing of several neighboring finite-rate modes, producing a narrow region of spectral compression. For the quadratically driven reservoir, by contrast, the gap continues to soften into the symmetry-broken regime while the neighboring finite-rate modes shift and spread toward larger decay rates. The gap mode and the surrounding finite-rate spectrum therefore become increasingly separated rather than softening collectively. The two transitions exhibit distinct spectral reorganizations, with localized compression near the first-order transition and growing gap--finite-rate separation across the continuous transition.

We next benchmark how these distinct spectral reorganizations relate to computational performance. Finite-size photon-density and gap scaling, including metastable and parity-sector interpretations, are presented in Sec.~I of the Supplemental Material~\cite{SM}.

\begin{figure}[htp]
    \centering
    \includegraphics[width=\columnwidth]{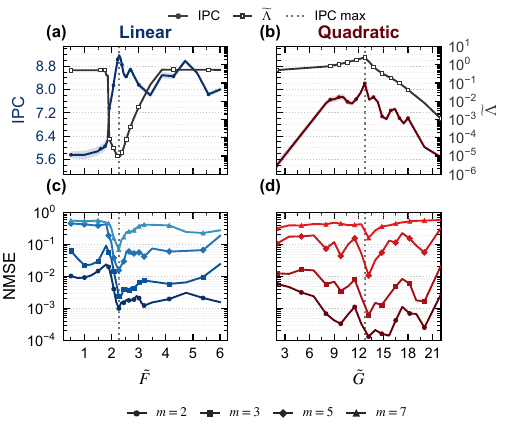}
    \caption{Information-processing performance across dissipative phase transitions. (a),(b) IPC and Liouvillian gap for the linearly and quadratically driven reservoirs. (c),(d) Normalized mean squared error (NMSE) for the NARMA-$m$ tasks, whose minima occur in the same parameter regions as the IPC maxima.
    }
    \label{fig:main_stmc_gap_narma}
\end{figure}

\textit{Enhanced information processing near dissipative phase transitions.—} 
We characterize temporal information processing using two complementary benchmarks, information processing capacity (IPC) and nonlinear NARMA prediction. The IPC quantifies the reconstruction of delayed Legendre-polynomial targets. For each task $\alpha=(d,\tau)$, the target is $y_\alpha(k)=P_d(2s_{k-\tau}-1)$, and the trained prediction is $\widehat y_\alpha(k) =W_{\rm out}^{(\alpha)}[\mathbf{x}_k,1]^T$. The corresponding capacity is $C_\alpha=r_\alpha^2$, with
\begin{equation}
    r_\alpha= \frac{ \left\langle \delta\widehat y_\alpha\,\delta y_\alpha \right\rangle_{\rm te}}{\sigma_{\widehat y_\alpha}\sigma_{y_\alpha}},
    \label{eq:main-task-correlation}
\end{equation}
where $\langle\cdot\rangle_{\rm te}$ denotes averaging over the test set, $\delta\widehat y_\alpha =\widehat y_\alpha-\langle\widehat y_\alpha\rangle_{\rm te}$ and $\delta y_\alpha=y_\alpha-\langle y_\alpha\rangle_{\rm te}$. The total information processing capacity is then
\begin{equation}
    C_{\rm IPC} =\sum_{d=1}^{D}\sum_{\tau=0}^{\tau_{\rm max}-1}C_{d,\tau}.
    \label{eq:main-stmc-total}
\end{equation}
Here the sum includes polynomial orders $d=1,2,3$ and delays $\tau=0,\ldots,9$, corresponding to $D=3$ and $\tau_{\rm max}=10$. To probe temporal processing that combines memory with nonlinear transformation, we use standard NARMA-$m$ prediction tasks and quantify the prediction error by the normalized mean squared error (NMSE). The task definitions and training procedure are given in the Supplemental Material~\cite{SM}.

Figure~\ref{fig:main_stmc_gap_narma} shows that temporal-processing performance is enhanced near both dissipative phase transitions, but with different relations to the Liouvillian gap. For the linearly driven reservoir, $C_{\rm IPC}$ reaches its maximum within the narrow first-order coexistence region, where the low-lying spectrum is compressed, as shown in Fig.~\ref{fig:spectral_reorganization}(a). The NARMA prediction errors are minimized in the same region. The agreement between these two benchmarks therefore identifies a robust performance optimum near first-order coexistence, which also coincides with the localized reorganization of the low-lying relaxation spectrum.

The quadratically driven reservoir shows a qualitatively different relation between performance and the relaxation spectrum. As $\tilde G$ increases from the symmetric phase, $C_{\rm IPC}$ reaches its maximum near the onset of symmetry breaking, and the NMSE values for the NARMA tasks exhibit corresponding minima. This optimum occurs where the Liouvillian gap begins to soften while the surrounding finite-rate spectrum starts to reorganize, as shown in Fig.~\ref{fig:spectral_reorganization}(b). Deeper in the symmetry-broken phase, the gap continues to soften, whereas the neighboring finite-rate modes shift and spread toward larger decay rates. Over the same parameter range, $C_{\rm IPC}$ decreases, while the NARMA prediction errors increase. The performance optimum therefore does not follow the continued softening of the Liouvillian gap, but remains near the onset of the gap--finite-rate spectral separation.

The IPC and NARMA optima establish enhancement across complementary benchmarks but do not identify the modes used by the trained reservoir. We resolve this attribution next; delay-resolved capacities are given in Sec.~III of the Supplemental Material~\cite{SM}.

\begin{figure}[htp]
    \centering
    \includegraphics[width=\columnwidth]{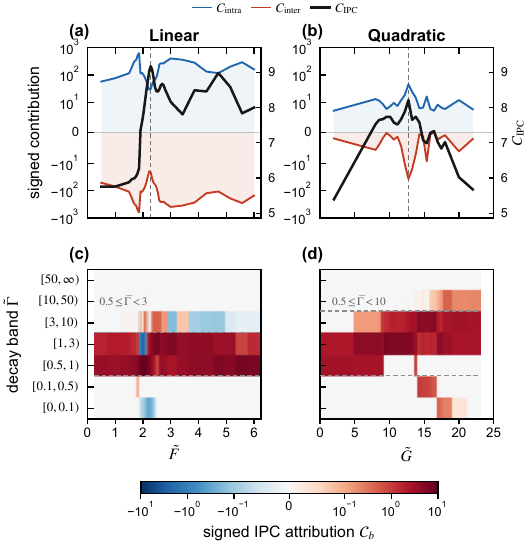}
    \caption{Liouvillian-mode attribution of the trained information processing capacity. (a) and (b) Intra-band term $C_{\rm intra}$, signed inter-band term $C_{\rm inter}$, and their sum $C_{\rm IPC}$ for the linearly and quadratically driven reservoirs. (c) and (d) Band-resolved signed attributions $\mathcal C_b=\sum_cK_{bc}$, grouped by the zero-input decay rate $\widetilde{\Gamma}=-\operatorname{Re}\lambda/\gamma$. }
    \label{fig:decomposition}
\end{figure}

\textit{Modal mechanisms of transition-enhanced performance.—}
Having established the information-processing optima near the two dissipative phase transitions, we next ask which intrinsic relaxation channels underlie the trained enhancement. The Liouvillian spectrum specifies the available relaxation timescales, but its eigenvalues alone do not determine whether a mode is excited by the input, visible to the measured observables, or assigned task-relevant weight by the trained readout. We therefore perform the modal attribution a posteriori in the biorthogonal eigenbasis of the zero-input Liouvillian. The resulting decomposition is exact and resolves the centered driven trajectory, measured features, and trained output in a common spectral basis; its derivation is given in the End Matter.

For each memory task $\alpha$, we group complex-conjugate Liouvillian modes into conjugate-closed decay-rate bands and write the centered trained prediction exactly as $\delta\widehat y_\alpha=\sum_bY_{b,\alpha}$. Denoting the normalized target covariance associated with band $b$ by $g_{b,\alpha}$, linearity of covariance gives $r_\alpha=\sum_b g_{b,\alpha}$. Introducing $K_{bc}=\sum_\alpha g_{b,\alpha}g_{c,\alpha}$ therefore yields
\begin{equation}
    C_{\rm IPC}=\sum_bK_{bb}+2\sum_{b<c}K_{bc}
    \equiv C_{\rm intra}+C_{\rm inter}.
    \label{eq:main-capacity-decomposition}
\end{equation}
Here $C_{\rm intra}$ collects the positive within-band terms, $C_{\rm inter}$ contains the signed cross-band contributions, and $\mathcal C_b=\sum_cK_{bc}$ gives the signed attribution associated with band $b$.

Figure~\ref{fig:decomposition} shows that the dominant trained attribution resides in finite-rate relaxation channels, whereas the slow sector containing the Liouvillian gap contributes only weakly. For the linearly driven reservoir, Fig.~\ref{fig:decomposition}(a) shows large positive $C_{\rm intra}$ and negative $C_{\rm inter}$ over much of the drive scan. Near the first-order transition, their magnitudes decrease together, weakening the destructive cross-band cancellation and allowing $C_{\rm IPC}$ to reach its maximum. More importantly, the band-resolved attribution in Fig.~\ref{fig:decomposition}(c) tracks the spectral reorganization in Fig.~\ref{fig:spectral_reorganization}(a). As the finite-rate modes soften near coexistence, the dominant positive attribution shifts from $1\leq\widetilde{\Gamma}<3$ to the slower $0.5\leq\widetilde{\Gamma}<1$ sector, and returns toward $1\leq\widetilde{\Gamma}<3$ away from coexistence. By contrast, the much slower $\widetilde{\Gamma}<0.5$ sector containing the gap remains weak near the optimum. The task-relevant attribution therefore tracks the collective softening of the finite-rate spectrum rather than the much stronger closing of the gap itself.

The quadratically driven reservoir exhibits the complementary evolution. Near the maximum of $C_{\rm IPC}$, Fig.~\ref{fig:decomposition}(b) shows an increase of the positive intra-band contribution that outweighs the accompanying negative inter-band term. The band-resolved attribution in Fig.~\ref{fig:decomposition}(d) again mirrors the intrinsic spectral reorganization in Fig.~\ref{fig:spectral_reorganization}(b). At smaller $\tilde G$, the dominant attribution is concentrated mainly in the slower finite-rate range $0.5\lesssim\widetilde{\Gamma}<3$. As $\tilde G$ increases and the neighboring Liouvillian modes shift and spread toward larger decay rates, the task-relevant attribution is redistributed toward $1\lesssim\widetilde{\Gamma}<10$, with the $3\leq\widetilde{\Gamma}<10$ sector becoming increasingly prominent. At the same time, the gap mode continues to soften into the slow sector and separates from these finite-rate channels, but its direct attribution remains comparatively small. Because the parity-odd input directly accesses this slow sector, the weak gap attribution cannot be explained simply by a symmetry-selection constraint. The decomposition thus makes explicit the separation already visible in the spectrum: the critical gap becomes progressively slower, whereas the trained information processing remains concentrated in finite-rate modes whose characteristic decay rates shift upward.

Taken together, Figs.~\ref{fig:spectral_reorganization} and \ref{fig:decomposition} establish a direct correspondence between transition-induced spectral reorganization and task-relevant modal attribution. In both cases, the gap-containing slow sector is only a weak contributor, while the dominant attribution follows the finite-rate relaxation channels and their signed cross-band combination. The enhanced performance is therefore associated with the reorganization of finite-rate relaxation dynamics rather than with the longest relaxation time itself. The robustness of these decay windows to the choice of band boundaries is examined in Sec.~IV of the Supplemental Material~\cite{SM}.

\textit{Experimental accessibility.—}
Experimental realizations closely related to the two reservoirs considered here are available in driven nonlinear resonators. A coherently driven superconducting Duffing oscillator has exhibited a first-order dissipative phase transition associated with long-lived metastable states~\cite{Chen20231stexp}, while both first-order and continuous symmetry-breaking transitions, including spontaneous $Z_2$ symmetry breaking, have been observed in a two-photon-driven superconducting Kerr resonator~\cite{beaulieu2025exp12}.

The finite-rate Liouvillian spectrum underlying this enhancement is likewise experimentally accessible through relaxation and correlation signals. For the same right and left Liouvillian eigenmodes used in the mode-resolved memory decomposition, a weak deviation $\delta\rho(0)$ from the stationary state evolves into a measured signal
\begin{equation}
\delta\langle O_j(t)\rangle=\sum_{\mu\neq0}A_{j\mu}\delta c_\mu(0)e^{\lambda_{\mu} t},
\label{eq:main-exp-relaxation}
\end{equation}
where $\delta c_{\mu}(0)=\langle\!\langle L_{\mu}|\delta\rho(0)\rangle\!\rangle$ is the excitation amplitude of mode $\mu$. Each Liouvillian eigenvalue can be written as $\lambda_\mu=-\Gamma_\mu+i\Omega_\mu$, where $\Gamma_\mu$ and $\Omega_\mu$ are the decay rate and oscillation frequency, respectively. Experimentally, these eigenvalues appear as poles of the Laplace-transformed relaxation or correlation signal,
\begin{equation}
\widetilde C_I(z) =\sum_{\mu\neq0} \frac{B_{I\mu}}{z-\lambda_\mu},
\label{eq:main-exp-poles}
\end{equation}
with $B_{I\mu}$ determined by the preparation and measurement overlaps. The decay-rate spectrum used above is therefore obtained from $\Gamma_\mu=-\operatorname{Re}\lambda_\mu$. Whenever a mode has nonzero preparation and measurement overlap, its relaxation rate and oscillation frequency can in principle be extracted from the measured time trace. The corresponding derivation and experimental precedents are summarized in the End Matter.

Our mode-resolved attribution shows that the enhanced memory is carried predominantly by finite-rate sectors rather than by the asymptotically slow gap mode. The measured spectrum alone does not determine reservoir performance, because the computational relevance of a mode also depends on its excitation by the input, visibility in the measured observables, and weighting by the trained readout. Once the task-relevant sectors are identified by the attribution, spectroscopy can directly test their reorganization across the transition and its relation to the observed performance enhancement. This provides a diagnostic analogous in spirit to the optical-response signature of a QRC memory sweet spot~\cite{qrcgenspectral}, but here directly resolves the Liouvillian relaxation channels associated with trained memory.

\textit{Discussion.—}
Previous studies have associated enhanced QRC performance with dynamical phase boundaries and quantum-chaotic regimes~\cite{PhysRevLettqrcphase,qrcphseedgeofchaos}, while a recent work connected memory performance to an experimentally accessible physical response~\cite{qrcgenspectral}. Here we go beyond such correspondences by establishing a mode-resolved quantitative attribution of the trained IPC directly to the intrinsic Liouvillian relaxation dynamics. This analysis shows that the computational enhancement is governed predominantly by finite-rate relaxation channels and by their signed cross-band combination, rather than by the asymptotically slow gap mode alone.

These results reveal a separation between the intrinsic relaxation structure of the reservoir and its task-dependent computational use. The autonomous Liouvillian spectrum defines the available relaxation channels, while the post-training modal attribution identifies those that contribute to the learned response. The phase transition reorganizes this relaxation structure, whereas the input protocol and trained readout determine how it is exploited for computation. This separation points to complementary routes for optimization at the level of both intrinsic relaxation dynamics and task-dependent mode utilization. Coherent and dissipative controls can reshape the Liouvillian relaxation landscape~\cite{Chen2022LEP,Zhou2023LEP,modecontrol2026}, while the input timescale can be tuned relative to the intrinsic relaxation times to favor task-relevant finite-rate channels and their constructive combination. These strategies can turn the spectral analysis developed here from a diagnostic tool into a physics-guided principle for designing dissipative quantum reservoirs. The same framework can also be extended to other diagonalizable Markovian reservoirs, including coupled driven-dissipative resonators~\cite{dudas2023qrcoscillator}.

\textit{Acknowledgments.—}
This work is supported by the National Natural Science Foundation of China (Grant No.12441502) and the Beijing Institute of Technology Research Fund Program under Grant No. 2024CX01015.

\paragraph*{Data Availability.—}
The data and numerical codes used to produce the reported results are available from the corresponding author upon reasonable request.

\nocite{Havel2003Lindblad,liovillianliang,macieszczak2016spectrual}
\bibliography{ref}

@article{Fujii2017,
  title = {Harnessing Disordered-Ensemble Quantum Dynamics for Machine Learning},
  author = {Fujii, Keisuke and Nakajima, Kohei},
  journal = {Phys. Rev. Appl.},
  volume = {8},
  issue = {2},
  pages = {024030},
  numpages = {20},
  year = {2017},
  month = {Aug},
  publisher = {American Physical Society},
  doi = {10.1103/PhysRevApplied.8.024030},
  url = {https://link.aps.org/doi/10.1103/PhysRevApplied.8.024030}
}

@article{Dambre2012IPC,
  author  = {Joni Dambre and David Verstraeten and Benjamin Schrauwen and Serge Massar},
  title   = {Information Processing Capacity of Dynamical Systems},
  journal = {Sci. Rep.},
  volume  = {2},
  pages   = {514},
  year    = {2012},
  doi     = {10.1038/srep00514}
}

@article{Ghosh2019QuantumReservoirProcessing,
  author  = {Sanjib Ghosh and Andrzej Opala and Micha{\l} Matuszewski and Tomasz Paterek and Timothy C. H. Liew},
  title   = {Quantum reservoir processing},
  journal = {npj Quantum Inf.},
  volume  = {5},
  pages   = {35},
  year    = {2019},
  doi     = {10.1038/s41534-019-0149-8}
}

@article{Drummond19801st,
  author  = {P. D. Drummond and D. F. Walls},
  title   = {Quantum theory of optical bistability. I. Nonlinear polarisability model},
  journal = {J. Phys. A},
  volume  = {13},
  pages   = {725--741},
  year    = {1980},
  doi     = {10.1088/0305-4470/13/2/034}
}

@article{Casteels171st,
  title = {Critical dynamical properties of a first-order dissipative phase transition},
  author = {Casteels, W. and Fazio, R. and Ciuti, C.},
  journal = {Phys. Rev. A},
  volume = {95},
  issue = {1},
  pages = {012128},
  numpages = {5},
  year = {2017},
  month = {Jan},
  publisher = {American Physical Society},
  doi = {10.1103/PhysRevA.95.012128},
  url = {https://link.aps.org/doi/10.1103/PhysRevA.95.012128}
}

@article{Chen20231stexp,
  author  = {Qi-Ming Chen and Michael Fischer and Yuki Nojiri and Michael Renger
             and Edwar Xie and Matti Partanen and Stefan Pogorzalek
             and Kirill G. Fedorov and Achim Marx and Frank Deppe
             and Rudolf Gross},
  title   = {Quantum behavior of the {Duffing} oscillator at the dissipative phase transition},
  journal = {Nat. Commun.},
  volume  = {14},
  pages   = {2896},
  year    = {2023},
  doi     = {10.1038/s41467-023-38217-x}
}

@article{shankar2013dissipativestate,
  title={Autonomously stabilized entanglement between two superconducting quantum bits},
  author={Shankar, Shyam and Hatridge, Michael and Leghtas, Zaki and Sliwa, KM and Narla, Aniruth and Vool, Uri and Girvin, Steven M and Frunzio, Luigi and Mirrahimi, Mazyar and Devoret, Michel H},
  journal={Nature},
  volume={504},
  number={7480},
  pages={419--422},
  year={2013},
  doi={10.1038/nature12802},
  publisher={Nature Publishing Group UK London}
}

@article{lin2013dissipativestate,
  title={Dissipative production of a maximally entangled steady state of two quantum bits},
  author={Lin, Yiheng and Gaebler, JP and Reiter, Florentin and Tan, Ting Rei and Bowler, Ryan and S{\o}rensen, AS and Leibfried, Dietrich and Wineland, David J},
  journal={Nature},
  volume={504},
  number={7480},
  pages={415--418},
  year={2013},
  doi={10.1038/nature12801},
  publisher={Nature Publishing Group UK London}
}

@article{barreiro2011quantumsimulation,
  title={An open-system quantum simulator with trapped ions},
  author={Barreiro, Julio T and M{\"u}ller, Markus and Schindler, Philipp and Nigg, Daniel and Monz, Thomas and Chwalla, Michael and Hennrich, Markus and Roos, Christian F and Zoller, Peter and Blatt, Rainer},
  journal={Nature},
  volume={470},
  number={7335},
  pages={486--491},
  year={2011},
  doi={10.1038/nature09801},
  publisher={Nature Publishing Group UK London}
}

@article{so2024dissimulation,
  title={Trapped-ion quantum simulation of electron transfer models with tunable dissipation},
  author={So, Visal and Duraisamy Suganthi, Midhuna and Menon, Abhishek and Zhu, Mingjian and Zhuravel, Roman and Pu, Han and Wolynes, Peter G and Onuchic, Jos{\'e} N and Pagano, Guido},
  journal={Sci. Adv.},
  volume={10},
  number={51},
  pages={eads8011},
  year={2024},
  doi={10.1126/sciadv.ads8011},
  publisher={American Association for the Advancement of Science}
}

@article{2018discorrection,
  title = {Coherent Oscillations inside a Quantum Manifold Stabilized by Dissipation},
  author = {Touzard, S. and Grimm, A. and Leghtas, Z. and Mundhada, S. O. and Reinhold, P. and Axline, C. and Reagor, M. and Chou, K. and Blumoff, J. and Sliwa, K. M. and Shankar, S. and Frunzio, L. and Schoelkopf, R. J. and Mirrahimi, M. and Devoret, M. H.},
  journal = {Phys. Rev. X},
  volume = {8},
  issue = {2},
  pages = {021005},
  numpages = {7},
  year = {2018},
  month = {Apr},
  publisher = {American Physical Society},
  doi = {10.1103/PhysRevX.8.021005},
  url = {https://link.aps.org/doi/10.1103/PhysRevX.8.021005}
}

@article{reiter2017dissipativecorrection,
  title={Dissipative quantum error correction and application to quantum sensing with trapped ions},
  author={Reiter, Florentin and S{\o}rensen, Anders S{\o}ndberg and Zoller, Peter and Muschik, Christine A},
  journal={Nat. Commun.},
  volume={8},
  number={1},
  pages={1822},
  year={2017},
  doi={10.1038/s41467-017-01895-5},
  publisher={Nature Publishing Group UK London}
}

@article{PRXQuantumqrcgenxun,
  title = {Quantum Reservoir Computing Using Arrays of Rydberg Atoms},
  author = {Bravo, Rodrigo Araiza and Najafi, Khadijeh and Gao, Xun and Yelin, Susanne F.},
  journal = {PRX Quantum},
  volume = {3},
  issue = {3},
  pages = {030325},
  numpages = {19},
  year = {2022},
  month = {Aug},
  publisher = {American Physical Society},
  doi = {10.1103/PRXQuantum.3.030325},
  url = {https://link.aps.org/doi/10.1103/PRXQuantum.3.030325}
}

@article{zia2025qrcgen,
  author  = {Danilo Zia and Luca Innocenti and Giorgio Minati
             and Salvatore Lorenzo and Alessia Suprano
             and Rosario Di Bartolo and Nicol{\`o} Spagnolo
             and Taira Giordani and Valeria Cimini
             and G. Massimo Palma and Alessandro Ferraro
             and Fabio Sciarrino and Mauro Paternostro},
  title   = {Quantum reservoir computing for photonic entanglement witnessing},
  journal = {Sci. Adv.},
  volume  = {11},
  number  = {50},
  pages   = {eady7987},
  year    = {2025},
  doi     = {10.1126/sciadv.ady7987}
}

@article{houqrcgen,
  title = {High-Accuracy Temporal Prediction via Experimental Quantum Reservoir Computing in Correlated Spins},
  author = {Hou, Yanjun and Hua, Juncheng and Wu, Ze and Xia, Wei and Chen, Yuquan and Li, Xiaopeng and Li, Zhaokai and Peng, Xinhua and Du, Jiangfeng},
  journal = {Phys. Rev. Lett.},
  volume = {136},
  issue = {12},
  pages = {120602},
  numpages = {8},
  year = {2026},
  month = {Mar},
  publisher = {American Physical Society},
  doi = {10.1103/r8ww-qw7j},
  url = {https://link.aps.org/doi/10.1103/r8ww-qw7j}
}

@article{PRXQuantumqrcgen,
  title = {Feedback-Driven Quantum Reservoir Computing for Time-Series Analysis},
  author = {Kobayashi, Kaito and Fujii, Keisuke and Yamamoto, Naoki},
  journal = {PRX Quantum},
  volume = {5},
  issue = {4},
  pages = {040325},
  numpages = {16},
  year = {2024},
  month = {Nov},
  publisher = {American Physical Society},
  doi = {10.1103/PRXQuantum.5.040325},
  url = {https://link.aps.org/doi/10.1103/PRXQuantum.5.040325}
}

@article{guoqrcgen,
  title = {Practical quantum reservoir computing in Rydberg atom arrays},
  author = {Liu, Dong-Sheng and Jie, Qing-Xuan and Zou, Chang-Ling and Ren, Xi-Feng and Guo, Guang-Can},
  journal = {Phys. Rev. A},
  volume = {113},
  issue = {4},
  pages = {042401},
  numpages = {9},
  year = {2026},
  month = {Apr},
  publisher = {American Physical Society},
  doi = {10.1103/pkhd-pl3w},
  url = {https://link.aps.org/doi/10.1103/pkhd-pl3w}
}

@article{tapioqrcgen,
  title = {Quantum reservoir computing on random regular graphs},
  author = {Ivaki, Moein N. and Lazarides, Achilleas and Ala-Nissila, Tapio},
  journal = {Phys. Rev. A},
  volume = {112},
  issue = {1},
  pages = {012622},
  numpages = {9},
  year = {2025},
  month = {Jul},
  publisher = {American Physical Society},
  doi = {10.1103/gq9r-d5q8},
  url = {https://link.aps.org/doi/10.1103/gq9r-d5q8}
}

@article{PhysRevLettqrcphase,
  title = {Dynamical Phase Transitions in Quantum Reservoir Computing},
  author = {Mart\'{\i}nez-Pe\~na, Rodrigo and Giorgi, Gian Luca and Nokkala, Johannes and Soriano, Miguel C. and Zambrini, Roberta},
  journal = {Phys. Rev. Lett.},
  volume = {127},
  issue = {10},
  pages = {100502},
  numpages = {7},
  year = {2021},
  month = {Aug},
  publisher = {American Physical Society},
  doi = {10.1103/PhysRevLett.127.100502},
  url = {https://link.aps.org/doi/10.1103/PhysRevLett.127.100502}
}

@article{our,
  title = {Robust and efficient quantum reservoir computing with a discrete time crystal},
  author = {Zhang, Da and Li, Xin and Guo, Yibin and Yu, Haifeng and Jin, Yirong and Yin, Zhang-Qi},
  journal = {Phys. Rev. Appl.},
  volume = {26},
  issue = {1},
  pages = {014056},
  numpages = {18},
  year = {2026},
  month = {Jul},
  publisher = {American Physical Society},
  doi = {10.1103/w2w9-c944},
  url = {https://link.aps.org/doi/10.1103/w2w9-c944}
}

@article{qrcphseedgeofchaos,
  title = {Edge of Many-Body Quantum Chaos in Quantum Reservoir Computing},
  author = {Kobayashi, Kaito and Motome, Yukitoshi},
  journal = {Phys. Rev. Lett.},
  volume = {136},
  issue = {4},
  pages = {040602},
  numpages = {8},
  year = {2026},
  month = {Jan},
  publisher = {American Physical Society},
  doi = {10.1103/j2qj-vwcl},
  url = {https://link.aps.org/doi/10.1103/j2qj-vwcl}
}

@article{qrcgenspectral,
  title = {Connection between Memory Performance and Optical Absorption in Quantum Reservoir Computing},
  author = {G\"otting, Niclas and Wilksen, Steffen and Steinhoff, Alexander and Lohof, Frederik and Gies, Christopher},
  journal = {Phys. Rev. Lett.},
  volume = {135},
  issue = {24},
  pages = {240403},
  numpages = {6},
  year = {2025},
  month = {Dec},
  publisher = {American Physical Society},
  doi = {10.1103/vp79-8t1l},
  url = {https://link.aps.org/doi/10.1103/vp79-8t1l}
}

@article{qrcnoise,
  author  = {L. Domingo and G. Carlo and F. Borondo},
  title   = {Taking advantage of noise in quantum reservoir computing},
  journal = {Sci. Rep.},
  volume  = {13},
  pages   = {8790},
  year    = {2023},
  doi     = {10.1038/s41598-023-35461-5}
}

@article{qrrqrcnoise23,
  title = {Temporal information processing induced by quantum noise},
  author = {Kubota, Tomoyuki and Suzuki, Yudai and Kobayashi, Shumpei and Tran, Quoc Hoan and Yamamoto, Naoki and Nakajima, Kohei},
  journal = {Phys. Rev. Res.},
  volume = {5},
  issue = {2},
  pages = {023057},
  numpages = {15},
  year = {2023},
  month = {Apr},
  publisher = {American Physical Society},
  doi = {10.1103/PhysRevResearch.5.023057},
  url = {https://link.aps.org/doi/10.1103/PhysRevResearch.5.023057}
}

@article{qrcdisdamp,
  author  = {Emanuele Ricci and Francesco Monzani and Luca Nigro and Enrico Prati},
  title   = {Quantum reservoir computing induced by controllable damping},
  journal = {npj Quantum Inf.},
  volume  = {12},
  pages   = {107},
  year    = {2026},
  doi     = {10.1038/s41534-026-01229-8}
}

@article{qrcdisearly,
  title = {Temporal Information Processing on Noisy Quantum Computers},
  author = {Chen, Jiayin and Nurdin, Hendra I. and Yamamoto, Naoki},
  journal = {Phys. Rev. Appl.},
  volume = {14},
  issue = {2},
  pages = {024065},
  numpages = {22},
  year = {2020},
  month = {Aug},
  publisher = {American Physical Society},
  doi = {10.1103/PhysRevApplied.14.024065},
  url = {https://link.aps.org/doi/10.1103/PhysRevApplied.14.024065}
}

@article{sannia2024dissipationqrc,
  author  = {Antonio Sannia and Rodrigo Mart{\'i}nez-Pe{\~n}a
             and Miguel C. Soriano and Gian Luca Giorgi
             and Roberta Zambrini},
  title   = {Dissipation as a resource for quantum reservoir computing},
  journal = {Quantum},
  volume  = {8},
  pages   = {1291},
  year    = {2024},
  doi     = {10.22331/q-2024-03-20-1291}
}

@article{qrcoscillator,
  title = {Quantum reservoir computing with a single nonlinear oscillator},
  author = {Govia, L. C. G. and Ribeill, G. J. and Rowlands, G. E. and Krovi, H. K. and Ohki, T. A.},
  journal = {Phys. Rev. Res.},
  volume = {3},
  issue = {1},
  pages = {013077},
  numpages = {9},
  year = {2021},
  month = {Jan},
  publisher = {American Physical Society},
  doi = {10.1103/PhysRevResearch.3.013077},
  url = {https://link.aps.org/doi/10.1103/PhysRevResearch.3.013077}
}

@article{dudas2023qrcoscillator,
  author  = {Dudas, Julien and Carles, Baptiste and Plouet, Erwan and Mizrahi, Frank Alice and Grollier, Julie and Markovi{\'c}, Danijela},
  title   = {Quantum reservoir computing implementation on coherently coupled quantum oscillators},
  journal = {npj Quantum Inf.},
  volume  = {9},
  pages   = {64},
  year    = {2023},
  doi     = {10.1038/s41534-023-00734-4}
}

@article{carles26qrccqed,
  title = {Experimental quantum reservoir computing with a circuit-quantum-electrodynamics system},
  author = {Carles, B. and Dudas, J. and Balembois, L. and Grollier, J. and Markovi\ifmmode \acute{c}\else \'{c}\fi{}, D.},
  journal = {Phys. Rev. Appl.},
  volume = {25},
  issue = {5},
  pages = {054005},
  numpages = {10},
  year = {2026},
  month = {May},
  publisher = {American Physical Society},
  doi = {10.1103/fmm7-3qm9},
  url = {https://link.aps.org/doi/10.1103/fmm7-3qm9}
}

@article{qrcoscijc,
  title = {Quantum reservoir computing in Jaynes-Cummings models: Nonlinear memory and time-series prediction},
  author = {Das, Sreetama and Giorgi, Gian Luca and Zambrini, Roberta},
  journal = {Phys. Rev. Res.},
  volume = {8},
  issue = {2},
  pages = {023148},
  numpages = {15},
  year = {2026},
  month = {May},
  publisher = {American Physical Society},
  doi = {10.1103/ffd3-ytbt},
  url = {https://link.aps.org/doi/10.1103/ffd3-ytbt}
}

@article{Xia2022qec,
  author  = {Wei Xia and Jie Zou and Xingze Qiu and Xiaopeng Li},
  title   = {The reservoir learning power across quantum many-body localization transition},
  journal = {Front. Phys.},
  volume  = {17},
  pages   = {33506},
  year    = {2022},
  doi     = {10.1007/s11467-022-1158-1}
}

@article{Martinezqrc2023Information,
  author  = {Rodrigo Mart{\'i}nez-Pe{\~n}a and Johannes Nokkala and Gian Luca Giorgi and Roberta Zambrini and Miguel C. Soriano},
  title   = {Information Processing Capacity of Spin-Based Quantum Reservoir Computing Systems},
  journal = {Cogn. Comput.},
  volume  = {15},
  pages   = {1440--1451},
  year    = {2023},
  doi     = {10.1007/s12559-020-09772-y}
}

@article{liovillianliang,
  title = {Symmetries and conserved quantities in Lindblad master equations},
  author = {Albert, Victor V. and Jiang, Liang},
  journal = {Phys. Rev. A},
  volume = {89},
  issue = {2},
  pages = {022118},
  numpages = {14},
  year = {2014},
  month = {Feb},
  publisher = {American Physical Society},
  doi = {10.1103/PhysRevA.89.022118},
  url = {https://link.aps.org/doi/10.1103/PhysRevA.89.022118}
}

@article{ciutispectral,
  title = {Spectral theory of Liouvillians for dissipative phase transitions},
  author = {Minganti, Fabrizio and Biella, Alberto and Bartolo, Nicola and Ciuti, Cristiano},
  journal = {Phys. Rev. A},
  volume = {98},
  issue = {4},
  pages = {042118},
  numpages = {13},
  year = {2018},
  month = {Oct},
  publisher = {American Physical Society},
  doi = {10.1103/PhysRevA.98.042118},
  url = {https://link.aps.org/doi/10.1103/PhysRevA.98.042118}
}

@article{disphaselukin,
  title = {Dissipative phase transition in a central spin system},
  author = {Kessler, E. M. and Giedke, G. and Imamoglu, A. and Yelin, S. F. and Lukin, M. D. and Cirac, J. I.},
  journal = {Phys. Rev. A},
  volume = {86},
  issue = {1},
  pages = {012116},
  numpages = {21},
  year = {2012},
  month = {Jul},
  publisher = {American Physical Society},
  doi = {10.1103/PhysRevA.86.012116},
  url = {https://link.aps.org/doi/10.1103/PhysRevA.86.012116}
}

@article{ciutiphase,
  title = {Exact steady state of a Kerr resonator with one- and two-photon driving and dissipation: Controllable Wigner-function multimodality and dissipative phase transitions},
  author = {Bartolo, Nicola and Minganti, Fabrizio and Casteels, Wim and Ciuti, Cristiano},
  journal = {Phys. Rev. A},
  volume = {94},
  issue = {3},
  pages = {033841},
  numpages = {11},
  year = {2016},
  month = {Sep},
  publisher = {American Physical Society},
  doi = {10.1103/PhysRevA.94.033841},
  url = {https://link.aps.org/doi/10.1103/PhysRevA.94.033841}
}

@article{macieszczak2016spectrual,
  title = {Towards a Theory of Metastability in Open Quantum Dynamics},
  author  = {Katarzyna Macieszczak and M{\u{a}}d{\u{a}}lin Gu{\c{t}}{\u{a}}
             and Igor Lesanovsky and Juan P. Garrahan},
  journal = {Phys. Rev. Lett.},
  volume = {116},
  issue = {24},
  pages = {240404},
  numpages = {6},
  year = {2016},
  month = {Jun},
  publisher = {American Physical Society},
  doi = {10.1103/PhysRevLett.116.240404},
  url = {https://link.aps.org/doi/10.1103/PhysRevLett.116.240404}
}

@article{mori2020resolvinggap,
  title = {Resolving a Discrepancy between Liouvillian Gap and Relaxation Time in Boundary-Dissipated Quantum Many-Body Systems},
  author = {Mori, Takashi and Shirai, Tatsuhiko},
  journal = {Phys. Rev. Lett.},
  volume = {125},
  issue = {23},
  pages = {230604},
  numpages = {6},
  year = {2020},
  month = {Dec},
  publisher = {American Physical Society},
  doi = {10.1103/PhysRevLett.125.230604},
  url = {https://link.aps.org/doi/10.1103/PhysRevLett.125.230604}
}

@article{modecontrol2026,
  title = {Relaxation Control of Open Quantum Systems},
  author = {Beato, Nicol\`o and Teza, Gianluca},
  journal = {Phys. Rev. Lett.},
  volume = {136},
  issue = {7},
  pages = {070401},
  numpages = {9},
  year = {2026},
  month = {Feb},
  publisher = {American Physical Society},
  doi = {10.1103/4frd-ck2z},
  url = {https://link.aps.org/doi/10.1103/4frd-ck2z}
}

@article{beaulieu2025exp12,
  author  = {Guillaume Beaulieu and Fabrizio Minganti and Simone Frasca
             and Vincenzo Savona and Simone Felicetti and Roberto Di Candia
             and Pasquale Scarlino},
  title   = {Observation of first- and second-order dissipative phase
             transitions in a two-photon driven {Kerr} resonator},
  journal = {Nat. Commun.},
  volume  = {16},
  pages   = {1954},
  year    = {2025},
  doi     = {10.1038/s41467-025-56830-w}
}

@article{Fink2018dptexp,
  author  = {Thomas Fink and Anne Schade and Sven H{\"o}fling
             and Christian Schneider and Ata{\c{c}} Imamoglu},
  title   = {Signatures of a dissipative phase transition in photon
             correlation measurements},
  journal = {Nat. Phys.},
  volume  = {14},
  pages   = {365--369},
  year    = {2018},
  doi     = {10.1038/s41567-017-0020-9}
}

@article{Havel2003Lindblad,
  author = {Timothy F. Havel},
  title = {Robust procedures for converting among {L}indblad, {K}raus and matrix representations of quantum dynamical semigroups},
  journal = {J. Math. Phys.},
  volume = {44},
  issue = {2},
  pages = {534--557},
  year = {2003},
  publisher = {AIP Publishing},
  doi = {10.1063/1.1518555}
}

@article{Wang2025Liouvillian,
  title = {Closed dynamical recursion equations for correlation functions and their application for the construction of the Liouvillian spectrum in Lindbladian systems},
  author = {Wang, Xueliang and Chen, Shu},
  journal = {Phys. Rev. B},
  volume = {111},
  issue = {9},
  pages = {094305},
  numpages = {28},
  year = {2025},
  month = {Mar},
  publisher = {American Physical Society},
  doi = {10.1103/PhysRevB.111.094305},
}

@article{Lax1963,
  title = {Formal Theory of Quantum Fluctuations from a Driven State},
  author = {Lax, Melvin},
  journal = {Phys. Rev.},
  volume = {129},
  pages = {2342--2348},
  year = {1963},
  doi = {10.1103/PhysRev.129.2342}
}

@article{Zhou2023LEP,
  title = {Accelerating relaxation through Liouvillian exceptional point},
  author = {Zhou, Yan-Li and Yu, Xiao-Die and Wu, Chun-Wang and Li, Xie-Qian and Zhang, Jie and Li, Weibin and Chen, Ping-Xing},
  journal = {Phys. Rev. Res.},
  volume = {5},
  issue = {4},
  pages = {043036},
  numpages = {13},
  year = {2023},
  month = {Oct},
  publisher = {American Physical Society},
  doi = {10.1103/PhysRevResearch.5.043036},
  url = {https://link.aps.org/doi/10.1103/PhysRevResearch.5.043036}
}

@article{Chen2022LEP,
  title = {Decoherence-Induced Exceptional Points in a Dissipative Superconducting Qubit},
  author = {Chen, Weijian and Abbasi, Maryam and Ha, Byung and Erdamar, Serra and Joglekar, Yogesh N. and Murch, Kater W.},
  journal = {Phys. Rev. Lett.},
  volume = {128},
  issue = {11},
  pages = {110402},
  numpages = {6},
  year = {2022},
  month = {Mar},
  publisher = {American Physical Society},
  doi = {10.1103/PhysRevLett.128.110402},
  url = {https://link.aps.org/doi/10.1103/PhysRevLett.128.110402}
}

@misc{SM,
  note = {See Supplemental Material for finite-size transition diagnostics and numerical conventions, benchmark and input-cadence details, delay-resolved information-processing capacity, and numerical validation and spectral-partition robustness of the Liouvillian-mode attribution.}
}

\section*{End Matter}
\textit{Exact Liouvillian-mode attribution.—} 
Because $\mathcal{L}_0$ is generally non-Hermitian, we use biorthogonal right and left eigenmodes $\{|R_{\mu}\rangle\rangle\}$ and $\{\langle\langle L_{\mu}|\}$ satisfying $\mathcal{L}_0|R_{\mu}\rangle\rangle=\lambda_{\mu}|R_{\mu}\rangle\rangle$, $\langle\langle L_{\mu}|\mathcal{L}_0=\lambda_{\mu}\langle\langle L_{\mu}|$ and $\langle\langle L_m|R_n\rangle\rangle=\delta_{mn}$. The memory capacities are evaluated from centered test sequences. At virtual node $v$, let $\bar\rho_v=\langle\rho_{\cdot,v}\rangle_{\rm te}$ and define
\begin{equation}
    |\delta\rho_{k,v}\rangle\rangle\equiv |\rho_{k,v}-\bar\rho_v\rangle\rangle
      =\sum_{\mu\neq0}\delta c_{\mu}(k,v)|R_{\mu}\rangle\rangle
    \label{eq:end-centered-state}
\end{equation}
with amplitude $\delta c_{\mu}(k,v)=\langle\langle L_{\mu}|\delta\rho_{k,v}\rangle\rangle$.
The stationary mode is absent because both $\rho_{k,v}$ and $\bar\rho_v$ have unit trace. The centered measured features therefore decompose as
\begin{equation}
    \delta x_{j,v}(k)
    =\sum_{\mu \neq0}A_{j\mu}\delta c_{\mu}(k,v),
    \label{eq:end-modal-feature}
\end{equation}
where $A_{j\mu}=\operatorname{Tr}(O_jR_{\mu})$ quantifies the visibility of mode $\mu$ through observable $O_j$.

For a memory task $\alpha=(d,\tau)$, the contribution of Liouvillian mode $\mu$ to the trained readout is
\begin{equation}
    y_{\mu,\alpha}(k)=\sum_{j,v}W_{jv}^{(\alpha)}A_{j\mu}\,\delta c_{\mu}(k,v).
    \label{eq:end-modal-waveform}
\end{equation}
The modal waveform is jointly determined by the input-induced excitation $\delta c_{\mu}$ and the mode-dependent readout factor $A_{j\mu}W_{jv}^{(\alpha)}$. Centering the trained prediction removes the fitted intercept and all time-independent feature offsets, so that the centered output is exactly recovered as $\delta\widehat y_\alpha(k)=\sum_{\mu \neq0}y_{\mu,\alpha}(k)$.

We next group complex-conjugate modes into conjugate-closed decay bands and define $Y_{b,\alpha}(k)=\sum_{\mu \in b}y_{\mu,\alpha}(k)$. For the centered target $\delta y_\alpha=y_\alpha-\langle y_\alpha\rangle_{\rm te}$, the normalized target covariance carried by band $b$ is
\begin{equation}
    g_{b,\alpha}
    =\frac{\langle Y_{b,\alpha}\,\delta y_\alpha\rangle_{\rm te}}
    {\sigma_{\widehat y_\alpha}\sigma_{y_\alpha}}.
    \label{eq:end-band-covariance}
\end{equation}
Linearity of covariance gives the Pearson coefficient of the complete prediction as $r_\alpha=\sum_b g_{b,\alpha}$. We collect the band-resolved contributions over all memory tasks in the symmetric kernel $K_{bc}=\sum_\alpha g_{b,\alpha}g_{c,\alpha}$. Defining $C_{\rm intra}=\sum_bK_{bb}$ and $C_{\rm inter}=2\sum_{b<c}K_{bc}$ recovers the capacity decomposition given in Eq.~\eqref{eq:main-capacity-decomposition}.
Finally, assigning half of each symmetric cross-band term to each participating band gives
\begin{equation}
    \mathcal C_b\equiv\sum_cK_{bc}=\sum_\alpha g_{b,\alpha}r_\alpha.
    \label{eq:end-band-attribution}
\end{equation}
Thus $\mathcal C_b$ is a signed attribution of the total trained capacity, not an isolated capacity of band $b$; negative inter-band terms correspond to target-correlated contributions with opposing signs. This exact identity underlies the intra-band, inter-band, and band-resolved quantities plotted in Fig.~\ref{fig:decomposition}.

\textit{Experimental extraction of the visible Liouvillian spectrum.—}
The relation used in the main text follows directly from the Liouvillian modal expansion. A small deviation from the steady state can be decomposed as
\begin{equation}
    \delta\rho(0)=\sum_{\mu \neq0}\delta c_{\mu}(0)R_{\mu} .
    \label{eq:end-exp-state}
\end{equation}
The modal amplitudes are $\delta c_{\mu}(0)=\langle\!\langle L_{\mu}|\delta\rho(0)\rangle\!\rangle$.After evolution under $\mathcal L_0$,
\begin{equation}
    \delta\rho(t)=\sum_{\mu\neq0}\delta c_{\mu}(0)e^{\lambda_{\mu} t}R_{\mu} .
    \label{eq:end-exp-evolution}
\end{equation}
Measurement through $O_j$ then gives Eq.~\eqref{eq:main-exp-relaxation}.A stationary correlation measurement provides a particularly direct experimental route to Liouvillian relaxation dynamics, as demonstrated by photon-correlation measurements of critical slowing near a dissipative phase transition~\cite{Fink2018dptexp}. Let $I(t)$ denote an experimentally recorded observable, such as a heterodyne quadrature or photon-number-related signal, and define its fluctuation by $\delta I=I-\langle I\rangle_{\rm ss}$. Its steady-state autocorrelation is
\begin{equation}
    C_I(t)=\big\langle\delta I(\tau+t)\,\delta I(\tau)\big\rangle_{\rm ss} .
    \label{eq:end-exp-correlation}
\end{equation}
Using the quantum regression theorem~\cite{Lax1963}, the initial operator perturbation associated with this correlation excites mode $\mu$ with amplitude
\begin{equation}
    \delta c_{\mu}^{(I)}=\langle\!\langle L_{\mu}|\delta I\rho_{\rm ss}\rangle\!\rangle .
    \label{eq:end-exp-corr-amplitude}
\end{equation}
The corresponding measurement visibility is $A_{I\mu}=\operatorname{Tr}(\delta I R_{\mu})$.

The measured correlation function therefore has the multimode form
\begin{equation}
    C_I(t)=\sum_{\mu\neq0}A_{I\mu}\,\delta c_{\mu}^{(I)}e^{\lambda_{\mu} t} .
    \label{eq:end-exp-correlation-modes}
\end{equation}
This expression makes clear that finite-time traces generally contain multiple visible Liouvillian modes. The familiar single-exponential extraction of the Liouvillian gap corresponds to the long-time limit in which the faster terms have already decayed.

The same information can be represented in the complex-frequency domain. Define the Laplace transform
\begin{equation}
    \widetilde C_I(z)=\int_0^\infty dt\,e^{-zt}C_I(t) .
    \label{eq:end-exp-laplace}
\end{equation}
Substitution of Eq.~\eqref{eq:end-exp-correlation-modes} gives
\begin{equation}
    \widetilde C_I(z)=\sum_{\mu\neq0}\frac{A_{I\mu}\,\delta c_{\mu}^{(I)}}{z-\lambda_{\mu}} .
    \label{eq:end-exp-poles}
\end{equation}
Hence each Liouvillian eigenvalue with nonzero preparation and measurement overlap appears as a pole of the measured correlation function~\cite{ciutispectral,Wang2025Liouvillian}. The experimentally visible decay rates are therefore obtained as $\Gamma_\mu=-\operatorname{Re}\lambda_\mu$ for modes with $A_{I\mu}\delta c_\mu^{(I)}\neq0$.

The relaxation rates entering our finite-rate spectral decomposition can therefore be extracted from experimentally recorded dynamical observables. By scanning the reservoir control parameter, one can reconstruct the evolution of the experimentally accessible relaxation spectrum and compare it with the corresponding memory performance. This provides a direct experimental test of the relation between transition-induced spectral reorganization and computational enhancement identified in the main text.

\end{document}